\documentclass[traditabstract]{aa}
\usepackage{natbib,twoopt}
\usepackage{graphicx}
\usepackage{txfonts}
\usepackage{color}
\usepackage{CJK}
\usepackage{caption}
\usepackage{subcaption}
\usepackage{dcolumn}
\usepackage{lscape}
\usepackage{amsmath}
\usepackage{booktabs}
\usepackage[colorlinks = true,linkcolor = blue,citecolor = blue]{hyperref}
\bibpunct{(}{)}{;}{a}{}{,}             
\makeatletter
  \newcommandtwoopt{\citeads}[3][][]{\href{http://adsabs.harvard.edu/abs/#3}%
    {\def\hyper@linkstart##1##2{}%
     \let\hyper@linkend\@empty\citealp[#1][#2]{#3}}}
  \newcommandtwoopt{\citepads}[3][][]{\href{http://adsabs.harvard.edu/abs/#3}%
    {\def\hyper@linkstart##1##2{}%
     \let\hyper@linkend\@empty\citep[#1][#2]{#3}}}
  \newcommandtwoopt{\citetads}[3][][]{\href{http://adsabs.harvard.edu/abs/#3}%
    {\def\hyper@linkstart##1##2{}%
     \let\hyper@linkend\@empty\citet[#1][#2]{#3}}}
  \newcommandtwoopt{\citeyearads}[3][][]%
    {\href{http://adsabs.harvard.edu/abs/#3}
    {\def\hyper@linkstart##1##2{}%
     \let\hyper@linkend\@empty\citeyear[#1][#2]{#3}}}
\makeatother

\begin{document}


\title{Chemical abundances in a high velocity RR Lyrae star near the bulge\thanks{The data presented herein were obtained at the W.M. Keck Observatory, which is operated as a scientific partnership among the California Institute of Technology, the University of California and the National Aeronautics and Space Administration. The Observatory was made possible by the generous financial support of the W.M. Keck Foundation. }
}

\titlerunning{Abundances in a high-velocity RR Lyrae star}

\author{C.J. Hansen \inst{1}, R.~M. Rich \inst{2}, A. Koch \inst{3,4}, S. Xu \inst{5}, A. Kunder\inst{6}, H.-G. Ludwig \inst{3}}
\authorrunning{C.~J. Hansen et al.}
\offprints{C.~J. Hansen, \email{cjhansen@dark-cosmology.dk}}

\institute{Dark Cosmology Centre, The Niels Bohr Institute, 
Juliane Maries Vej 30, DK-2100 Copenhagen, Denmark
\email{cjhansen@dark-cosmology.dk}
\and
University of California Los Angeles, Department of Physics \& Astronomy, Los Angeles, CA, USA
\and
Zentrum f\"ur Astronomie der Universit\"at Heidelberg,  Landessternwarte, K\"onigstuhl 12, 69117 Heidelberg, Germany
\and
Phyics Department, Lancaster University, Lancaster LA1 4YB, UK
\and
European Southern Observatory, Karl-Schwarzschild-Stra{\ss}e 2, 85748, Garching, Germany
\and
Leibniz-Institut f\"ur Astrophysik Potsdam, An der Sternwarte 16, D-14482 Potsdam, Germany
}
\date{Received February 2, 2016; accepted March 16, 2016}

\abstract{Low-mass, variable, high-velocity stars are interesting study cases for many aspects of Galactic structure and evolution. 
Until recently, the only known high- or hyper-velocity stars were young stars thought to originate from the Galactic centre. Wide-area surveys like APOGEE and BRAVA have found several low-mass stars in the bulge with Galactic rest-frame velocities larger than 350 km/s. In this study we present the first abundance analysis
of a low-mass, RR Lyrae star, located close to the Galactic bulge, with a space motion of $\sim -400$\,km/s. 
Using medium-resolution spectra, we derive abundances (including upper limits) of 11 elements. 
These allow us to chemically tag the star and discuss its origin, although 
our derived abundances and metallicity, at [Fe/H] $=-0.9$ dex,  do not point toward one unambiguous answer. 
Based on the chemical tagging, we cannot exclude that it originated in the bulge. 
However, combining its retrograde orbit and the derived abundances suggests that the star was accelerated  from the outskirts
of the inner (or even outer) halo during many-body interactions. 
Other possible origins include the bulge itself, or the star could be stripped from a star cluster or the Sagittarius dwarf galaxy when it merged with the Milky Way.}
\maketitle

\section{Introduction}
RR Lyrae stars are short-period variables that are luminous, old Population II stars, and therefore ideal for numerous astronomical studies. Owing to their pulsations (with periods of 0.2--1\,days) we can determine their distances via the period-luminosity relation, while their high luminosity allows us to detect them to greater distances in remote systems compared to what is possible with fainter, less evolved dwarf and giant stars.

RR Lyrae stars have very compact centres containing most of their mass while very little mass is found in the outer 50\% (in radius) of the star \citep{Smith1995}. The RR Lyrae lie in the instability strip on the horizontal branch and thus have temperatures between 6100 - 7400\,K. Most RR Lyrae stars are on the zero age horizontal branch (ZAHB) which means they are burning He to C in the core. In addition to the core burning H is fused to He in a shell. The exact chemical composition, especially the He content, will dictate the evolution and pulsation properties of these variable stars \citep{Smith1995}. Depending on how the RR Lyrae pulsate (fundamental note versus first overtone) they are grouped in three subclasses (Bailey types); RRab vs RRc where the former shows asymmetric light curves with larger variations in magnitude compared to the latter c type which show more smooth low amplitude variations. According to \citet{Smith1995,Preston2011} the typical pulsational velocity of RRab are $60-70$\,km/s (however, larger amplitudes can be reached depending on the star's V magnitude \citealt{Liu1991}) while those of RRc only vary with $30-40$\,km/s.  Finally, the multi-mode RRd variables pulsate in both the fundamental and the first overtone and they compose the smallest subgroup among these three types \citep{Brown2004}.
 Previous studies \citep[e.g.,][]{Clementini1995,For2011,Hansen2011b,Haschke2012} have shown that despite the evolved stage of RR Lyrae stars, they seem to preserve the original surface composition they were born with for most of the heavy elements (except for lithium). This makes them excellent study cases for chemical tagging in the Milky Way as well as in Local Group galaxies \citep[e.g.,][]{Brown2004,DaCosta2010}.

Their age and low metallicity of the RR Lyrae stars make them useful probes of the original chemical composition of the old components of the Milky Way, namely the halo and bulge.   \citet{Layden1996} showed, using a kinematic study, that some of the more metal-rich ([Fe/H] $=-1.0$) RR Lyrae stars belong to the thick disk. This was confirmed by the more recent Northern Sky Variability Survey \citep{Kinemuchi2006} despite their uncertainties in visual amplitudes and thus V magnitudes. An indication of the Galactic component in which the RR Lyrae could reside is given by the stellar metallicity. A [Fe/H]$>-1.0$ might indicate that the star belongs to the thick disk or bulge, while a lower value $-1.0$ to $-2.5$ would point towards it being an (inner/outer) halo star.
When looking towards the central regions of the Milky Way, metallicity alone becomes an insufficient parameter and large surveys have found an overlap of the bulge component with the metal-weak thick disk and the inner halo \citep[i.e., within 3.5 kpc, e.g.,][]{fulbright2007,Kunder2012,Ness2013,Johnson2014}. 
This distinction is significant, since the inside-out formation of the Galaxy dictates that the oldest and most (extremely) metal-poor stars should be found in the very center \citep{Tumlinson2010}, rendering them inner halo stars (by
formation and chemistry) that happen to be located within the bulge. In fact, while still low in numbers, the very metal-poor ``bulge'' stars known to date show chemical abundances that vastly overlap with those 
of the metal-poor halo distribution \citep[e.g.,][]{Casey2015,KochMcW2016,Howes2015}. Their kinematics then indicates that their orbits are mainly confined to the inner few kpc of the Galaxy (although the star we consider here is not confined to the innermost parts).   

For this reason the high-velocity star (HVS) investigated in \citet{Kunder2015} is of particular interest, since it turned out to be a metal-poor RR Lyrae (type ab) star moving with a high Galactic rest-frame velocity of $-482$ km/s on an eccentric orbit around the bulge and into the halo. 
Previously, most of the high-to-hyper-velocity stars were thought to be young (relatively massive) hot stars originating from the Galactic centre \citep[e.g.,][ and references therein]{Hills1988,Brown2015}. However, the recent SEGUE study by \citet{Palladino2014} used a combination of radial velocities and proper motions to show that low-mass F and G stars could be traveling with high enough speed to allow them to escape the Galaxy, and that these stars did not originate in the Galactic Centre \citep[see also][]{Kollmeier2010,Nidever2012}. \citet{Geier2015} confirmed these findings by conducting a kinematic and spectroscopic follow-up of the fastest known HVS star which travels at $\sim 1200$\,km/s through the Galaxy. Moreover, this star (\object{US~708}, a hot subdwarf), was found to be a low-mass, compact He star that on top of its fast space motion is also a fast rotator  \citep[][and references therein]{Geier2015}. 
The recent study by \citet{Li2015} found 19 HVS candidate stars in  the first LAMOST data release, and they suggest that they may originate from the Galactic bulge. The work by Kunder et al. 2016 (in press) also finds a subgroup of high velocity RR lyrae stars in the bulge.

This makes the very evolved RR Lyrae star of  \citet{Kunder2015} an interesting study case, as it is to our knowledge the first high-velocity evolved RR Lyrae star to travel at high speed {\em between} different Galactic components (disk, bulge and halo). 
Despite its current position close to the bulge, it is believed that the RR Lyrae star (MACHO 176.18833.411) originates in the halo, a conclusion reached by tracing back  the calculated orbits. Here we conduct a detailed medium-resolution follow-up study of this star to trace and understand the origin of this high-velocity RR Lyrae star. 

\section{Observations and data reduction}
\begin{figure}[!ht]
\begin{center}
\includegraphics[width=0.49\textwidth]{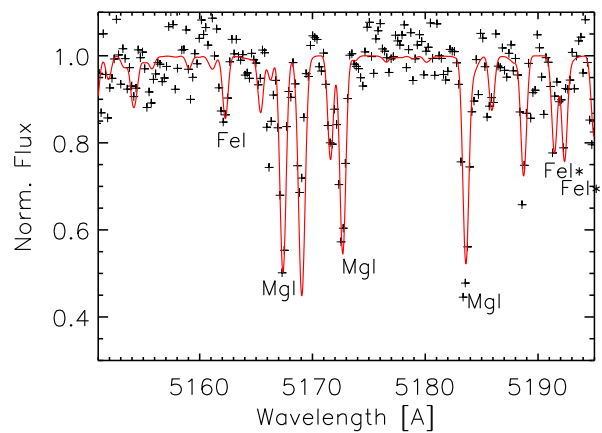}
\caption{The spectral region of m176 around the Mg triplet. Overall the fit is acceptable, however, we are biased towards measuring strong lines which do not always yield the same abundance as seen from weaker lines. A few of the weaker, yet blended, Fe lines are shown here and marked with a `*' (the two redmost lines).}
\label{fig:obs}
\end{center}
\end{figure}

\object{MACHO 176.18833.411} (hereafter m176) was observed three times during 24 April 2015 using the Echellette Spectrograph and Imager (ESI; \citealt{Sheinis2002}
) at Keck II with a slit of 0.5" resulting in a resolution of $R\sim 8000$. Each exposure lasted 1800\,s and was obtained at phase 0.55 -- 0.63 which coincides with the minimum light phase (for further details see Table~\ref{tab:obslog}). This star has a stable period of 0.5152d according to \citet{OGLE2014} and \citet{Kunder2015}. With an integration time of 1800\,s, the observations last about 4\% of the pulsation period. This is approximately a factor of three longer than the integration times used in \citet{For2011}, however, our integration times are still short enough to avoid smearing of the spectra and the lines maintain a clean Gaussian profile (see Fig.~\ref{fig:obs}). This star is fairly faint ($<V>$=17.586), which makes it hard to obtain a high-resolution, high-S/N spectrum. The OGLE V and I light curves, as shown in \citet{Kunder2015}, clearly show the characteristic asymmetric variations of an RR Lyrae ab type variable.

The data were reduced using the Makee pipeline\footnote{Makee was developed by T. A. Barlow specifically for reduction of Keck HIRES data. 
It is freely available online at the Keck Observatory home page, \tt{http://www.astro.caltech.edu/$\sim$tb/makee/}.}, which conducts standard data
reduction operatios such as bias subtraction, flatfielding, wavecalibration including
several wavelength corrections, such as applying air-to-vacuum and
heliocentric velocity corrections. Since the spectra were obtained almost at the same
phase, the extracted 1D spectra could be co-added to a single spectrum to increase the
signal-to-noise ratio (SNR). Following the spectrum was shifted to rest
wavelength and the continuum was normalised in IRAF by dividing the spectrum
with a fitted pseudo continuum. The final spectrum spans a wavelength range from $\sim 3800 - 11000$\,\AA.
A SNR of $\sim 40$ per pixel (or $\sim 130$ per \AA) was estimated from a spectral region around 7000\,\AA. 

\begin{table*}
\centering
\caption{Date of observation, coordinates, heliocentric Julian date (HJD), integration times, and phase for m176.}
\label{tab:obslog} 
\begin{tabular}{lccccc}
\hline
Date &Ra  &  Dec & HJD & Exp. time & phase\\
 & & & &[s] &\\
\hline
04.24.2015 & 18:00:17.35 & -27:18:07.4 &2457137.07022& 1800 & 0.545\\
04.24.2015 & 18:00:17.35 & -27:18:07.4 &2457137.09172& 1800 & 0.587\\
04.24.2015 & 18:00:17.35 & -27:18:07.4 &2457137.11321& 1800 & 0.628\\
\hline
\end{tabular}
\end{table*}

\section{Stellar parameters and abundances}
The stellar parameters were derived using only the spectra. As a first pass, we measured equivalent widths (EWs) for 17 Fe I and four Fe II lines. These lines allowed for a first rough determination of the stellar parameters. At this resolution, most Fe lines are blended or too noisy to be useful for determining the parameters (see Fig.~\ref{fig:obs}), and we ended up with 13 Fe I and two Fe II lines (see Table~A\ref{tab:lines}). 
The remaining 15 useful Fe lines were used to determine an excitation temperature by requiring that all Fe I lines yield the same Fe abundance regardless of excitation potential. The microturbulence velocity was set by requiring that all Fe lines produce the same Fe abundance, and the gravity was fixed by changing this model parameter until the same abundance values for Fe I and Fe II were obtained. Since we only have two Fe II lines, this parameter is more uncertain with respect to the other parameters that rely on Fe I measurements. The value obtained for log$g$ is however reasonable for RR Lyrae stars. The metallicity was updated by synthesising weaker Fe lines, which yielded a somewhat lower [Fe/H] value than derived from the strong EW measurements. The final set of stellar parameters is (T [K]/log$g$/[Fe/H]/$\xi_{mic}$ [km/s]): $6600\pm100/2.0\pm0.3/-0.9\pm0.2/4.5\pm0.2$.

In order to obtain an improved estimate of  the gravity, we use the Yale isochrones \citep{Demarque2004} and an IDL program by \citet{Yong2013}. However, the program fails to return a value for this very evolved RR Lyrae star.
The [Fe/H] from weak line synthesis was confirmed by measuring the bluest Ca line (8498\,\AA) in the near-infrared Ca II triplet and inserting the EW into the [Fe/H]-EW(Ca) relation of \citet{Gomez2011,Wallerstein2012}. Several independent EW measurements yielded a metallicity of [Fe/H] $=-0.85\pm 0.15$ which is in excellent agreement with the weak line value of [Fe/H] $=-0.9$. The uncertainty in the Ca EW originates from continuum placement and to a lesser extent from the profile fitted (Gauss versus Voigt -- note that a Lorentzian should not be used as it overestimates the EW and in turn the [Fe/H] value).

\begin{figure*}[!ht]
\begin{center}
\includegraphics[width=0.96\textwidth]{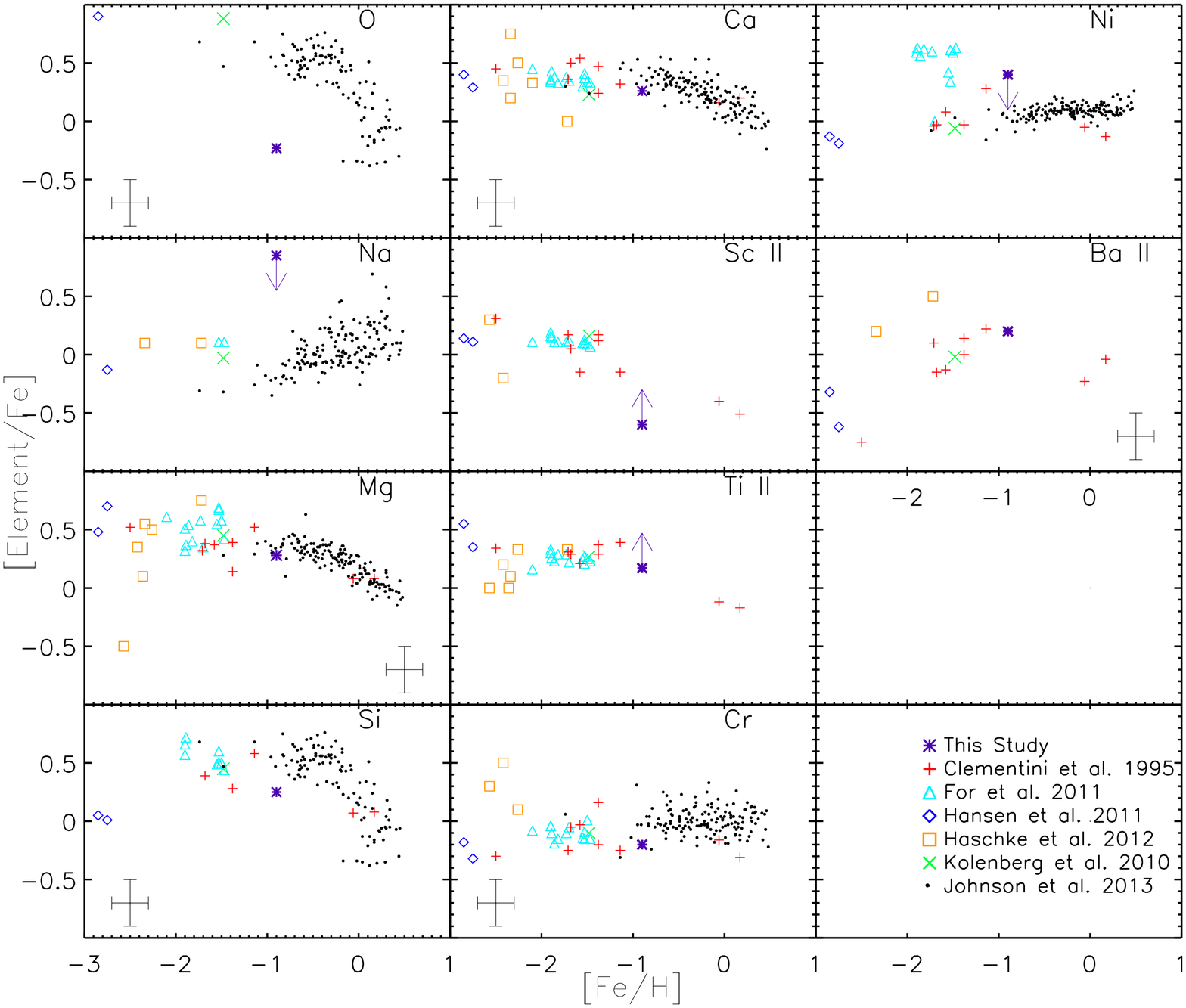}
\caption{Abundances of m176 compared to other RR Lyrae stars observed in a similar phase (around minimum light) as well as bulge stars. The RR lyrae star studies shown are: \citet{Clementini1995,For2011,Hansen2011b,Kolenberg2010,Haschke2012} and \citet{Johnson2013} as bulge comparison sample. \label{fig:abundances}}
\end{center}
\end{figure*}

The stellar abundances were derived using an interpolated new-ODF (new opacity distribution function) ATLAS9 model atmosphere \citep{Castelli2003,Allende2004} in conjunction with MOOG spectrum synthesis code \citep[][version 2014]{Sneden1973}. We decided to only rely on spectral lines in the region 5000 -- 9000\,\AA~ in order to get the cleanest lines (with fewest blends and lower noise) to derive the best possible abundances. This, on the other hand, also means that we are using medium to strong lines for this analysis, since the spectrum quality prevents us from using weak lines. By conducting spectrum synthesis we determine abundances (including upper limits) for eleven elements which are listed in Table~\ref{tab:results} (see also Fig.~\ref{fig:abundances}).
The lines to synthesise were selected from the line list in \citet{Hansen2011b} in order to find the lines most likely detectable in RR Lyrae stars. The synthesis was conducted using a line list containing atomic data from \citet{Sneden2014, Hansen2013, Bergemann2012, Gallagher2012,Ivans2006, Lawler2001}.

\begin{table}
\centering
\caption{Element, stellar abundances and line-to-line dispersion, and number of lines. A $^{wa}$ indicates that a weighted average is listed, where one of the three Cr lines is given weight 0.5.}
\label{tab:results} 
\begin{tabular}{lccc}
\hline
Element, X &  [X/Fe]  &$\sigma$ & No. lines\\
\hline
O I  & $-0.23$& $\pm0.1$ & 3\\
Na I& $<0.85$ & -- & 2 \\
Mg I& 0.28  &$\pm0.11$ & 4\\
Si I& 0.25  & -- & 1 \\
Ca I& 0.26  & $\pm0.1$ & 6 \\
Sc II&$>-0.6$ & -- & 1 \\
Ti II & $<0.17 $ & -- & 4 \\
Cr I & $-0.2^{wa}$ & $\pm0.1$ & 2.5\\
Ni I & $<0.4$ &  -- & 1 \\
Ba II & 0.2& $\pm0.0$ & 2\\
\hline
\end{tabular}
\end{table}

From Table~\ref{tab:results} and Fig.~\ref{fig:abundances} the derived abundances and upper limits for m176 are seen to be in good agreement with values derived for other RR Lyrae stars. Comparing m176 to the bulge study by \citet{Johnson2013} the abundances generally could agree with those derived for bulge stars, however, the level of $\alpha-$enrichment also agrees well with values published for the halo \citep{Cayrel2004,Nissen2010}. The Cr and Si abundances of m176 are slightly lower than what is reported for the bulge stars \citep{Johnson2013}. For the Cr abundance we derived values for three lines, but one of them was more noisy and yielded a different value, which is why we only assigned this line half weight when calculating the final Cr abundance (hence the 2.5 lines listed in Table~\ref{tab:results}). The only real outlier is the oxygen abundance of m176. We used the O-triplet lines at $7774-7777$\,\AA, which are affected by NLTE and 3D effects \citep{Caffau2008, Nissen2014, Steffen2015, Amarsi2016}. Both corrections would need to be computed exactly for this star. However, for the Sun the corrected O$_{trip}$ abundances would increase by $\sim0.15$\,dex \citep{Steffen2015}, so if we corrected our O abundance it would most likely still be lower than the abundances derived for the other $\alpha-$elements. 

The abundances seem to indicate that this RR Lyrae star is a rather normal halo star at [Fe/H] $=-0.9$ placing it in the more metal-rich tail of the halo metallicity distribution \citep{Schoerck2009}. However, the metallicity also matches the moderately metal-poor stars in the bulge, where \citet{Li2015} recently found 19 candicate HVS could originate from; for more details see Sect.~\ref{results} and \ref{halo}. 
The lines analysed and their atomic data are presented in the online appendix (Table~A\ref{tab:lines}).

\subsection*{Uncertainties}
The abundance uncertainties arise from stellar parameters, continuum placement, and the quality of the spectral fit which, owing to the SNR and the medium resolution, is of slightly lower precision compared to high-resolution, higher SNR spectra. This results in a line-to-line abundance variation of up to 0.2\,dex, as seen for Fe and $<$0.11\,dex for Mg when using lines weaker than the Mg triplet lines. (Fig.~\ref{fig:obs} shows that the strong Mg-triplet lines are difficult to fit with one single abundance.)
The stellar parameter uncertainties of $\pm100$K/$\pm0.3$/$\pm0.2$/$\pm0.2$\,km/s result in an abundance uncertainty of $\sim 0.13 - 0.2$\,dex depending on the element. The uncertainty related to the continuum placement is typically 0.05 and for the spectrum fit it varies from higher SNR to lower SNR regions from 0.02 -- 0.1\,dex. By adding all the uncertainties in quadrature 
we obtain a total uncertainty of 0.16 -- 0.22\,dex. For simplicity we adopt an average value of 0.2\,dex for all elements in the figures.

\subsection*{Helium}
The He~I line at 5875.6\,\AA~ is an interesting absorption feature as it would tell us about the helium abundance in the star. The amount of helium is a governing factor in the pulsation and evolution of very evolved stars (such as RR Lyrae stars). The layer in which helium is being doubly ionised is responsible for absorbing heat during the stellar compression, and, in turn, for driving and maintaining the pulsations in the so-called $\gamma$-mechanism.
However, the helium line is only visible in the maximum light phase (0.9-1.0), and since we have observed this RR Lyrae at minimum light, this line is unfortunately not detectable. Moreover, He I and II emission lines are also visible in the rising light phases or close to the peak, and the line strengths depend on the pulsation period, velocity amplitude, and metal abundance (for more details on He line behaviour see \citealt{Preston2011}).

\subsection{Impact of microturbulence}
Compared to other, less evolved stars the RR Lyrae stars have high microturbulences,  $\xi_{mic}$. Moreover, the medium resolution and a SNR$\sim 40$ prevents us from using weak lines, hence we suspect a few of the strongest lines to be close to saturation. A higher microturbulent velocity could therefore help desaturate the lines and this would in turn provide more reliable abundances.
We therefore test the impact of using Kurucz models with new opacity distributions, created with microturbulences of 2, 4, and 8\,km/s. The changes are  not detectable for most of the lines we analysed. 
A few of the stronger, and more sensitive, lines show changes of around 0.05\,dex, and in those cases we produce lines where the Gaussian synthetic line profiles are better broadened and therefore fit the observed spectrum better. However, thermal broadening remains most important at these temperatures (which are also somewhat higher in RR Lyrae than in dwarfs and giants).  
Only the strongest lines like the Na D lines show very large differences when synthesized with a microturbulence of 8\,km/s compared to 4\,km/s. These lines are affected by the outer regions of the stellar atmospheres, where the T-p (temperature-pressure) profiles of the different models  (in $\xi_{mic}$) differ the most. The Na D lines were difficult to fit, and possibly saturated, so we do not give this result too much weight, which is why our [Na/Fe] is an upper limit.

\section{Results and Discussion \label{results}}
In order to understand the origin and enrichment of this high velocity RR Lyrae star we must first analyse and interpret our derived surface abundances. 
Evolved horizontal branch and RR Lyrae stars have proven to be trustworthy study cases in galactic archeology studies (e.g., \citealt{Tautvaisiene1997,Preston2006} as well as the two most metal-poor RR lyrae stars studied \citealt{Hansen2011b}). 
However, when evolved and expanded giant stars move with high velocities through the ambient interstellar medium 
 they might become subject to mass loss, if the outer layers are not strongly bound (this is, e.g., the case for the luminous, low-gravity, variable Mira stars, where material from the outer layers are being removed by ram pressure stripping, e.g., \citealt{Martin2007}). Given the compactness of RR Lyrae stars, this is 
 unlikely to be of a concern even in fast-moving ones.  
But before drawing any conclusions based on the abundances we now check if pulsations combined with the high velocity can lead to mass loss from the stellar surface.

\subsection*{Escape velocity}
Is material being stripped from this HVS RR Lyrae star or does it stay bound? As mentioned in the introduction, the outermost 50\% by radius of a typical RR Lyrae star contains $\sim$0.001 M$_{\odot}$ 
which could in principle be stripped from the star.
If we assume a typical RR Lyrae star mass of 0.65--0.8\,M$_{\odot}$ and a radius of 4--6\,R$_{\odot}$, we adopt a mass of 0.7 \,M$_{\odot}$ and a radius of 5\,R$_{\odot}$ \citep{Smith1995,Catelan2015}\footnote{With these or slightly lower values we would get a log$g$ of $\sim2.5$ which is slightly higher than what we derive from the spectra. This is an expression of the few strong lines we had to rely on and the bias of LTE Fe abundance.}. 
These allow us to calculate the escape velocity from the stellar surface using:\\
\begin{displaymath}
V_{escape} = \sqrt{2\,G\,M/R} = 231.2\,{\rm km/s}
\end{displaymath}

The expansion velocity of the outer layers, in turn,  can be deduced from the change in velocity along the line of sight from phase 0 to 0.5 which corresponds to the change from minimum radius to maximum radius. 
According to the \citet{Liu1991} pulsational velocity amplitude vs V light amplitude relations, a velocity of $\sim90$\,km/s was estimated based on m176's V-band amplitude of 1.27 mag (see Fig. 1 in \citeauthor{Liu1991}). 
Thus for m176, this value is V$_{expansion} \sim 90$\,km/s \citep[see Fig.~1 in][]{Kunder2015}. 
This means that V$_{expansion} <<$ V$_{escape}$ and the surface material will stay bound owing to the relative compactness of this class of stars (as confirmed by earlier RR Lyrae star abundance studies). 
This confirms that the surface abundances reflect the original chemical composition and have not been stripped or altered owing to the combination of the pulsational and high spatial velocity of the star.

\subsection*{High Velocity Stars}
As mentioned in the introduction, the first HVSs to be detected were massive, hot stars that originated from the Galactic centre  \citep{Hills1988}. Recent studies \citep[e.g.,][]{Palladino2014} have shown that low-mass F or G stars can also travel with extremely high velocities, often exceeding the local Galactic escape velocity. 
Thus, stars in a cluster or a binary system can be kicked out e.g., by a perturbation of the system, such as in a three-body encounter or a supernova explosion \citep[e.g.,][]{Hoogerwerf2000,Perets2009,Irrgang2010,Zhang2013}.
To be more specific, there are several possible formation scenarios for HVSs: 1) interaction of single-stars with a central massive black hole (MBH), 2) tidal break-up of binary stars in the vicinity of a MBH, 3) three-body interactions involving 
single-star encounters with a binary or cluster of MBHs, 4) double detonation of SN Ia with a close system of a rapidly orbiting low-mass compact He star and a massive ($\sim 1-1.2M_{\odot}$) CO-white dwarf \citep{Geier2015}, 5) single degenerate SN type Ia consisting of a white dwarf and main sequence star \citep{Liu2013}, 6) binary star disruption in dense, interacting regions like globular clusters or the Galactic bulge/disk, or 7) tidal disruption of dwarf galaxies in the Galaxy \citep[][and references therein]{Li2015}. The last case would lead to old rapidly traveling stars. 

In all the scenarios involving a binary system, we would expect some pollution or mass transfer to have taken place. This could possibly change the original surface composition by enhancing the He, carbon and possibly the neutron capture element abundances. If the star were to be kicked out of a cluster some slow neutron-capture pollution might be expected.
We could derive neither molecular nor atomic C from the medium-resolution spectra as the star is warm (6600K) and the blue spectra are very noisy, but our spectra would still have allowed us to 
detect significant C-enhancement. Moreover, the [Ba/Fe] abundance is low, and when comparing this to Fig. 5 in \citet{Stancliffe2013} the mass transfer, if it occurred, would have been much lower than 0.1\,M$_{\odot}$. Thus  the surface composition of m176\footnote{This star may be part of the high velocity dispersion subgroup of RR Lyrae stars in the bulge (Kunder et al. 2016, subm.).}  would only have been affected by a negligible amount in the measurable element abundances.

\subsection{Metallicity: Light curve vs. spectroscopic measurements}
All of our spectroscopic measurements are in excellent internal agreement with each other: based on a representative number of 16 Fe I lines we found [Fe/H] =$-0.90\pm0.2$ dex, which compares to the well-calibrated 
Ca triplet measurement of [Fe/H]$_{\rm CaT}=-0.85\pm0.15$ dex. In contrast, the photometrically derived metallicity values suggest this star is considerably more metal-poor.  A Fourier decomposition of the OGLE I-band light curve and the \citet{Smolec2005} Fourier-[Fe/H] calibration yields a photometric metallicity of $-$1.26 dex on the \citet{Jurcsik1996} metallicity scale, or $-$1.52 dex on the \citet{Carretta2009} metallicity scale.  \citet{Kunder2015} performed a Fourier decomposition on the OGLE V-band light curve and used the widely adopted \citet{Jurcsik1996} Fourier-[Fe/H] calibration to find a photometric metallicity of $-$1.62 dex on the \citet{Carretta2009} metallicity scale with a deviation parameter\footnote{$D_m$ is defined as 
$|F_{\rm obs}-F_{\rm calc}|/\sigma$, where F$_{\rm obs}$ is the observed value for any given Fourier parameter, F$_{\rm calc}$ is the predicted value based on the remaining parameters, and 
$\sigma$ is the observed standard deviation in $F$.} 
$D_m$ = 3.6.  \citet{Jurcsik1996} cautioned that relating physical stellar properties to Fourier parameters is applicable and "reliable" only if $D_m$ < 3, although some studies have relaxed this criteria to $D_m$ < 5 \citep[e.g.,][]{Cacciari2005}.  The disagreement between the spectroscopic and photometric [Fe/H] may indicate that careful consideration of the deviation parameters of (bulge) RR Lyrae stars needs to be taken, which is a limitation when using the \citet{Smolec2005} calibration, as it does not provide a way to calculate a deviation parameter and hence excludes such a `sanity' check.  Ultimately, a larger sample of stars with well determined spectroscopic metallicities is needed to clarify the discrepancy between our spectroscopic and the photometrically determined [Fe/H].  

The accuracy of photometric metallicities derived from the light curve Fourier parameters is not well established for the RR Lyrae population in the direction of the bulge.
We note that ours is the first medium-resolution spectroscopic  study of a 
RR Lyrae star in the direction of the Galactic bulge.  \citet{Kunder2008} find that a star-to-star comparison between photometric metallicity of bulge RR Lyrae stars in Baade's Window and those from low-resolution spectroscopic metallicities of \citet{Walker1991} have a dispersion twice as large as what would be expected from the \citet{Jurcsik1996} calibrating sample.  In contrast, the photometric [Fe/H] determinations of LMC stars compared to that of low-resolution spectroscopy agree to within 0.2 dex. It has also been shown that Fourier components of the light curves of RR Lyrae stars in the metal-rich bulge clusters \object{NGC~6411} and \object{NGC~6388} give photometric metallicity values that are ~0.6 dex more metal-poor than what is spectroscopically observed \citep{Sandage2004}.  This may indicate that RR Lyrae stars toward the direction of the bulge have properties distinct from the majority of local field RR Lyrae stars that make up the photometric [Fe/H] calibrating sample (e.g., different $\alpha-$abundances, helium abundances, or evolutionary channel), rendering the photometric [Fe/H] values for these stars uncertain.

\subsection{Chemical inferences on the birthplace of m176 \label{halo}}
Using the Besan\c con model of the Milky Way \citep{Robin2003}, we find that no halo RR Lyrae star more metal-rich than $-1.3$\,dex should be found towards this direction toward the bulge.
As is known, from stellar population synthesis, it is not as common for RR Lyrae to be produced from metal-rich systems, so the probability of metal-rich RR Lyrae forming in a metal-poor system 
like the halo is low \citep{Lee1992,Layden1995}. 
So if this star is not a typical bulge object, but a halo star as indicated by kinematics, then the  
 Galactic halo model implies that an RR Lyrae star with a relatively high metallicity so close to the bulge suggests
 that   it could have an extragalactic origin.

The orbits presented in \citet{Kunder2015} seem to indicate that the star (m176) would originate from the halo and is presently passing through the bulge. 
However, we do not know if it came from the outer or inner halo, or if, as the authors suggest, the star is less likely to originate from the bulge. A low or high $\alpha-$abundance allows us to distinguish between the outer and inner halo  \citep{Nissen2010} population. The inner halo shows a metallicity distribution function peaking around $<$[Fe/H]$> \sim-1.6$ with stars on prograde orbits that are thought to reflect the old(est) population of the Galaxy, while the outer halo is found to have an even lower mean metallicity with stars on retrograde orbits \citep{Carollo2007}. The latter could point towards stars being stripped from dwarf galaxies as a result of a galaxy merger where smaller systems were absorbed into the Galaxy. In either case, the outer halo is nowadays believed to be built up from smaller subsystems \citep{Bullock2005}, which would have limited gas resources, thus a different initial mass function for systems with a lower mass than that of the Milky Way. This would, in turn, lead to lower $\alpha-$abundances \citep[e.g.,][]{Kobayashi2006,Nissen2010}. 
Similarly, the low star forming efficiencies of the dwarf galaxy satellites in question lead to a downturn in the [$\alpha$/Fe] ratios already at low [Fe/H], again leaving the impression of strong $\alpha$-depletions 
towards higher metallicities \citep[e.g.,][]{Matteucci1990, Tolstoy2009, Hendricks2014}. 
Hence, a low [Mg/Fe] ($\sim0.2$\,dex) in a star on retrograde orbits could point towards an external origin of a star that is now located in the outer halo, while a high ($\sim0.4$) [Mg/Fe] in a star on a prograde orbit would indicate an old population formed in situ (inner halo).

\begin{figure}[!ht]
\begin{center}
\includegraphics[width=0.49\textwidth]{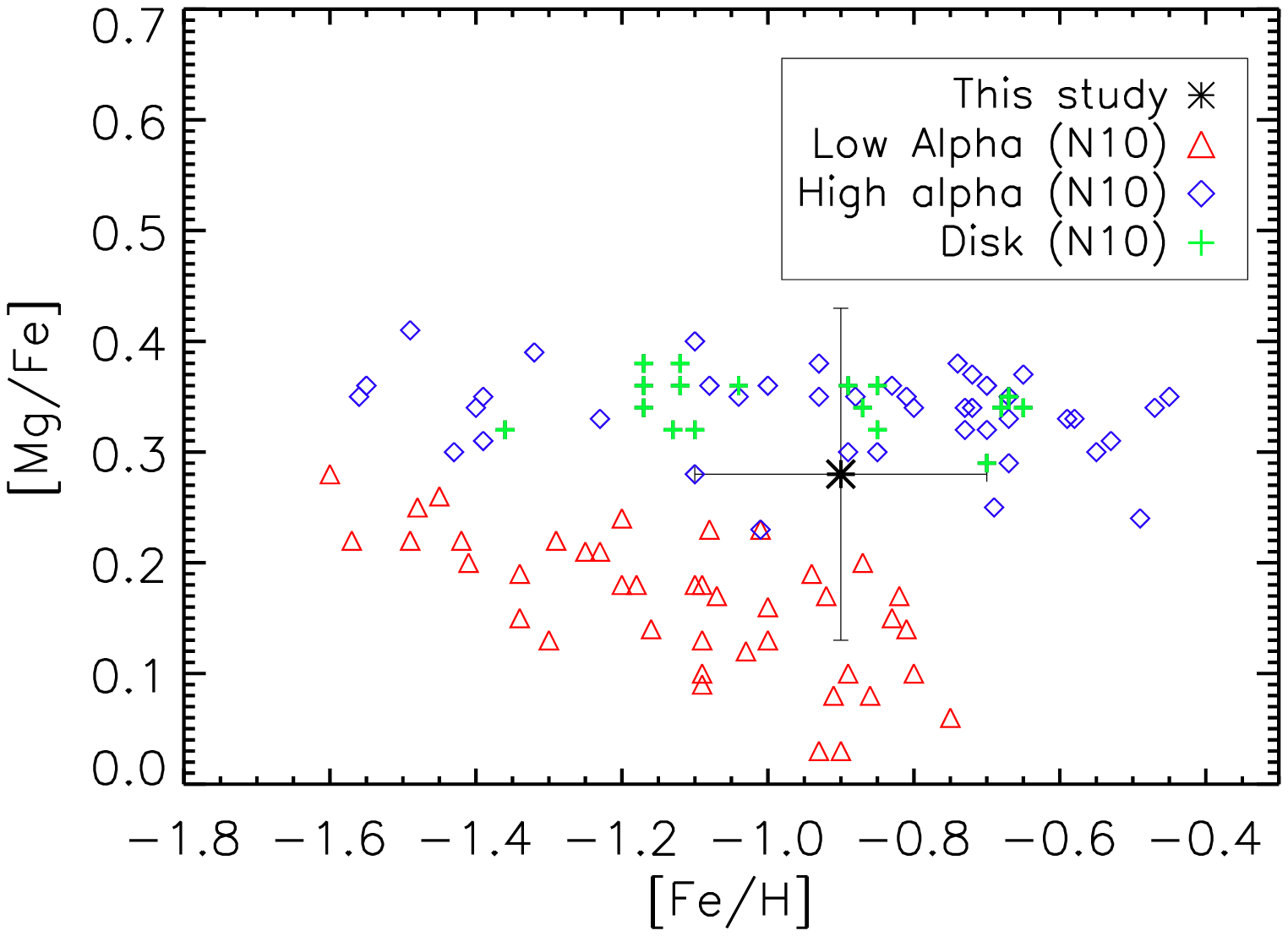}
\includegraphics[width=0.49\textwidth]{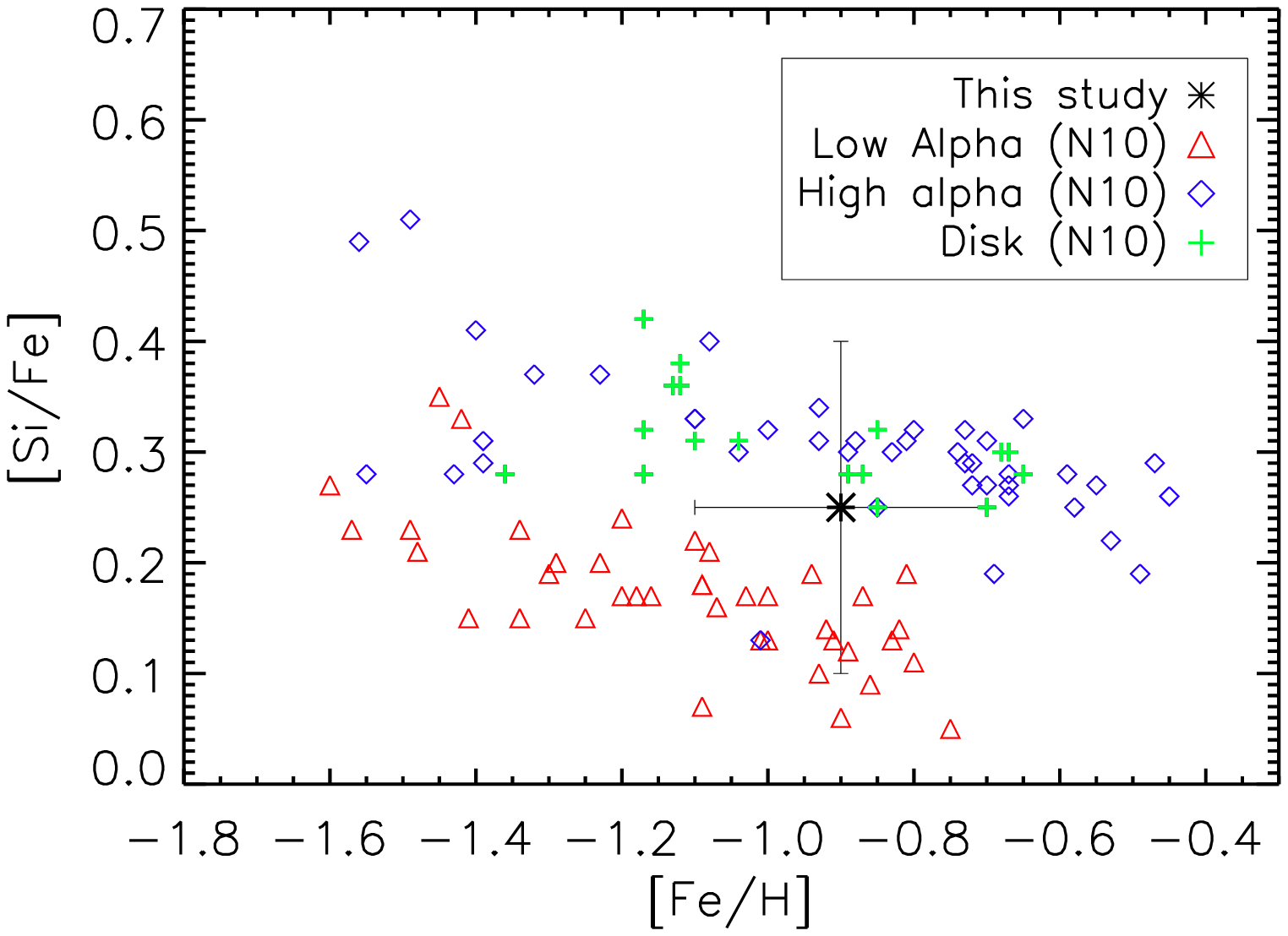}
\caption{Abundances of m176 compared to the high/low $\alpha$ abundances in (inner/outer) halo and thick disk stars from \citet{Nissen2010}.}
\label{fig:nissen}
\end{center}
\end{figure}

\citet{Nissen2010} presented a large sample of stars from the thick disk, inner and outer halo, all observed at high resolution. We compare the abundances of our RR Lyrae star (m176), which we showed to represent the original birth composition and remain unmodified by the stellar evolution and high velocity, to the study of \citet{Nissen2010}.
Figure~\ref{fig:nissen} shows this comparison for [Mg/Fe] and [Si/Fe].

\begin{figure}[!ht]
\begin{center}
\includegraphics[width=0.49\textwidth]{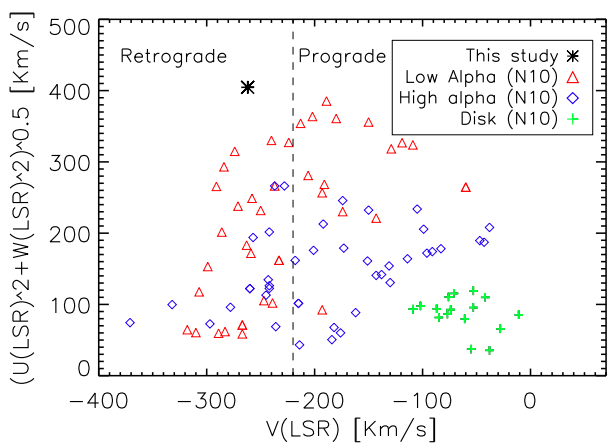}
\caption{Local Standard of rest velocities of m176 compared to the halo and thick disk stars from \citet[][N10]{Nissen2010}.}
\label{fig:toomre}
\end{center}
\end{figure}

This RR Lyrae star lies right between the inner halo and thick disk stars on one hand and the outer halo stars on the other hand. In the [Mg/Fe] panel of Fig.~\ref{fig:nissen} the RR Lyrae star (m176) might be slightly closer to the inner halo/thick disk stars. 
In \citep{Nissen2014} C and O abundances were presented for the same sample as published in their 2010 paper. The oxygen abundance we find fall $\sim0.4$\,dex below their lowest outer halo O abundances.
Hence, we need to consider the stellar kinematics
to determine whether the star is moving on an prograde or retrograde orbit. 
As in \citet{Nissen2010} we construct a Toomre diagram, where we also compare to their data. The result is shown in Fig.~\ref{fig:toomre}, from which it is clearly seen that m176 moves on a retrograde orbit that is even more extreme than the stars from \citet{Nissen2010}. This indicates that the high velocity RR Lyrae star, m176, originated from a not too remote part of the outer halo (this is also feasible compared to the orbit predictions provided in \citealt{Kunder2015}). 
The fact that the star moves with such a high velocity could indicate that it was originally kicked out from a binary system or that it is a result of a galaxy merger event, where a dwarf galaxy would be incorporated into the Milky Way and some of the dwarf's stars would be stripped in the process. This could explain the kinematics and velocity of the RR Lyrae star m176.  

According to \citet{deBoer2015} the first stripping from the Sagittarius (Sgr) dwarf galaxy resulted in stream stars with a mean metallicity of $\sim -1.5$ around 11--13 Gyr ago, while a second, more metal-rich stream ($\sim -0.7$) is younger ($\sim5$Gyr; see also \citet{Fellhauer2006}). 
The fact that RR Lyrae are generally old Pop II stars is consistent with them being stripped from a dwarf galaxy (e.g., Sgr) in an early merger event
 since the age ($>10$Gyr) would allow the star to have been stripped early on. In addition to this, the earlier study by \citet{Monaco2007} showed that there is a southern and a northern stream which peak around $<[$Fe/H$]> = -0.61$ and  $<[$Fe/H$]> = -0.83$, respectively. 
 
 In addition, they also found an increasing $\alpha-$abundance trend with decreasing metallicity for both the Sgr stream and main body. Their [Mg/Fe] values span values from $\sim -0.3$ to $0.4$\,dex. By combining the fact that RR Lyrae stars are thought to belong to an old population, that the Sgr stream consists of older and younger streams peaking at different mean metallicities close to the [Fe/H] $=-0.9$ we obtained for m176, and that positive (almost halo-like) [Mg/Fe] values can occur; the stripping from Sgr option remains a viable origin of m176. 
In fact, an extragalactic origin has been proposed for peculiar HVSs in the Milky Way (\citet{Gualandris2007}; but cf. \citealt{Brown2010}). 

Lastly, we consider that m176 is a ``normal'' RR Lyrae star located towards the direction of the bulge, which is on the tail of the RR Lyrae star velocity distribution.  Its spectroscopic [Fe/H] is what is expected for a RR Lyrae stars located toward the bulge \citep[e.g.,][]{ Walker1991,Pietrukowicz2015} as is its distance from the Galactic center.  Although it has a halo-like orbit, it may well be that the majority of ``bulge'' RR Lyrae stars reside in an inner halo \citep[e.g.,][]{Minniti1999,Kunder2008}.  This is also suggested by their spatial distribution (\citealt{Dekany2013}, although see also \citealt{Pietrukowicz2015}).  A larger sample of spectroscopically well-studied bulge RR Lyrae stars, as well as a larger sample of bulge RR Lyrae stars with space velocities is needed to address the uniqueness of this star.

\section{Conclusion}
The high velocity RR Lyrae star, m176, seems at first glance to be a normal halo star from a chemical perspective  with a slight $\alpha-$enhancement, a normal Cr and Ba abundance and upper limits that agree well with our understanding of the halo stars and their chemical imprints. It is remarkable to find a variable star from an old population with a space velocity of $-$482\,km/s with respect to the Galactic rest frame. It has completed many orbites around the bulge during its lifetime, hence repeated passages through/near the bulge, and yet, despite variability and high velocity, it preserved the original birth composition for the elements studied here.

The $\alpha-$abundances are in good agreement except for O which needs to be consolidated from high-resolution follow-up spectroscopy. However, all $\alpha-$abundances are $\leq0.3$dex and combined with the LSR velocity this indicates that the star is likely a halo star on a retrograde orbit, which associates it with the outer halo. Thus, this star could through stellar interaction originate from a binary system that interacted with a massive object, a binary system residing in a stellar cluster at the verge of the inner/outer halo region, or from dense parts of the bulge where interactions are more frequent. In all cases, a higher density will lead to more frequent interactions as well as a higher binary fraction.                                                                                                                                                                              
The normal Ba abundance and the lack of C seems to go against this explanation in terms of a binary origin. However, based on the metallicity and $\alpha-$abundances we cannot 
determine at this point  if the star was stripped from a system in the halo or the bulge. Interactions with a binary system, stellar cluster or a dense region in either of these two Galactic components could result in a HVS. Given the metallicity of $\sim -1 $ and the currently growing  population of HVS RR Lyrae stars near the bulge -- this would speak in favour of a bulge origin.
Another possible explanation for the origin of m176 is that it was tidally stripped from a different galaxy (a dwarf or an early merger event such as Sgr) which is supported by the age and the stellar abundances (low O and Ti). 

Higher resolution spectra of this star as well as other low-mass HVS stars would allow us to perform a more detailed abundance analysis, including weaker lines, of this and similar stars so that we can look deeper into their formation and origin. This will be important to understand if there is a population of (HVS) RR lyrae stars in the bulge region.

\begin{acknowledgements}
C.J.H. acknowledges support from research grant VKR023371 from the
Villum Foundation. We thank the anonymous referee for comments, and G. Wallerstein, R. Stancliffe and A. Ruiter for useful discussions. R.M.R acknowledges support from the NSF grant  AST-1413755.
A.K. thanks the Deutsche Forschungsgemeinschaft for funding from  Emmy-Noether grant  Ko 4161/1. H.G.L. acknowledges financial
support by the Sonderforschungsbereich SFB 881 "The Milky Way System" (subproject A4) of the German Research Foundation (DFG).
The authors wish to recognize and acknowledge the very significant cultural role and reverence that the summit of Mauna Kea has always had within the indigenous Hawaiian community.  We are most fortunate to have the opportunity to conduct observations from this mountain. 
\end{acknowledgements}

\bibliographystyle{aa}
\bibliography{/Users/cjhansen/Master_bib}

\Online
\appendix

\begin{table}
\centering
\caption{\label{tab:lines} Atomic data for lines investigated: Wavelength, atomic number,
ionisation, and isotope, excitation potential, and log$gf$. For the lines with
hyper fine structure (HFS) the total log$(gf)$ has been listed. Scandium
has been split according online data from Kurucz's hyperfine line list, barium according to \citet{Gallagher2012},
Europium according to \citet{Lawler2001,Ivans2006}. }

\begin{tabular}{lccr}
\hline
$\lambda$ [\AA] &  $Z$   & Ex.pot. [eV] & log$gf$\\
\hline
  7771.944 &      8.0  &     9.139   &     0.320   \\ 
  7774.166 &      8.0  &     9.139   &     0.170   \\ 
  7775.388 &      8.0  &     9.139   &    $-0.050$   \\ 
  5889.951 &     11.0  &     0.000   &     0.120   \\ 
  5895.924 &     11.0  &     0.000   &    $-0.180$   \\ 
  5167.321 &     12.0  &     2.707   &    $-1.030$   \\ 
  5172.684 &     12.0  &     2.710   &    $-0.400$   \\ 
  5183.604 &     12.0  &     2.715   &    $-0.180$   \\ 
  5528.405 &     12.0  &     4.343   &    $-0.620$   \\ 
  6155.134 &     14.0  &     5.615   &    $-0.400$   \\ 
  5588.749 &     20.0  &     2.524   &     0.210   \\ 
  5857.451 &     20.0  &     2.930   &     0.230   \\ 
  6102.723 &     20.0  &     1.878   &    $-0.890$   \\ 
  6122.217 &     20.0  &     1.884   &    $-0.410$   \\ 
  6439.075 &     20.0  &     2.524   &     0.470   \\ 
  6493.781 &     20.0  &     2.519   &     0.140   \\ 
  5526.79$^{HFS}$ &  21.1  &   1.767   &   0.020  \\ 
  5186.325 &     22.0  &     2.115   &    $-1.050$   \\ 
  5186.847 &     22.0  &     3.543   &    $-2.270$   \\ 
  5226.543 &     22.1  &     1.565   &    $-1.300$  \\ 
  5418.768 &     22.1  &     1.581   &    $-2.130$  \\ 
  5204.510 &     24.0  &     0.941   &    $-0.190$   \\ 
  5206.040 &     24.0  &     0.941   &     0.020   \\ 
  5208.420 &     24.0  &     0.941   &     0.170   \\ 
  5509.103 &     24.0  &     4.609   &    $-1.370$   \\ 
  5509.910 &     24.0  &     4.452   &    $-1.140$   \\ 
  5508.606 &     24.1  &     4.153   &    $-2.110$  \\ 
  5510.702 &     24.1  &     3.824   &    $-2.452$  \\ 
  4920.503 &     26.0  &     2.830   &     0.068  \\ 
  4966.087 &     26.0  &     3.332   &    $-0.871$  \\ 
  5339.928 &     26.0  &     3.266   &    $-0.647$  \\ 
  5405.775 &     26.0  &     0.990   &    $-1.844$  \\ 
  5501.465 &     26.0  &     0.958   &    $-3.046$  \\ 
  5506.779 &     26.0  &     0.990   &    $-2.789$  \\ 
  6065.482 &     26.0  &     2.609   &    $-1.410$  \\ 
  6136.995 &     26.0  &     2.198   &    $-2.950$  \\ 
  6137.692 &     26.0  &     2.588   &    $-1.403$  \\ 
  6393.601 &     26.0  &     2.433   &    $-1.576$  \\ 
  6430.846 &     26.0  &     2.176   &    $-1.946$  \\ 
  6494.980 &     26.0  &     2.404   &    $-1.239$  \\ 
  6677.987 &     26.0  &     2.692   &    $-1.418$  \\ 
  5316.615 &     26.1  &     3.153   &    $-1.870$  \\ 
  5534.847 &     26.1  &     3.245   &    $-2.865$  \\ 
  5476.904 &     28.0  &     1.825   &    $-0.780$  \\ 
  6643.638 &	 28.0  &     1.676   &	  $-2.300$  \\
  5853.67$^{HFS}$&  56.1  &  0.604   &    $-1.010$   \\ 
  6496.90$^{HFS}$&  56.1  &  0.604   &    $-0.380$   \\
  6645.11$^{HFS}$&  63.1  &  1.379   &     0.120  \\
\hline
\end{tabular}
\end{table}

\end{document}